
%
%
\def\drho{{\partial _\rho}}

\def\drhobar{{\partial _{\bar \rho}}}
\def\dsr{{{{\partial^2 \rho}\over{\partial z^2}}}}
\def\half {{1 \over 2}}
\def\dz{{\partial _z}}
\def\dzbar{{\partial _{\bar z}}}
\def\dy{{\partial _y}}

\def\va{{v^\alpha}}
\def\vb{{v^\beta}}
\def\vba{{\bar v^\alpha}}
\def\vbb{{\bar v^\beta}}
\def\m {{m^\mu}}
\def\plb{{+\bar l}}

\def\pmb{{+\bar m}}
\def\pnb{{+\bar n}}

\def\mm{{- m}}

\def\ml{{- l}}
\def\hp{{h^+}}
\def\hm{{h^-}}

\def\etp {{\eta^+}}
\def\etm {{\eta^-}}
\def\gmu {{\gamma^\mu_{\alpha\beta}}}

\def\vep {{\varepsilon^+}}
\def\vem {{\varepsilon^-}}

\def\xp {{\xi^+}}
\def\xm {{\xi^-}}

\def\php {{\phi^+}}
\def\phm {{\phi^-}}

\def\sp {{\psi^+}}
\def\sm {{\psi^-}}
\def\sbp {{\bar\psi^+}}
\def\sbm {{\bar\psi^-}}
\def\dzxp {{\dz x^{9+0} +\half\sp\dz\sm+\half\sm\dz\sp}}

\def\xplb {{x^\plb}}
\def\xml {{x^\ml}}
\def\Gplb {{\Gamma^\plb}}
\def\Gml {{\Gamma^\ml}}
\def\Gbplb {{\bar\Gamma^\plb}}
\def\Gbml {{\bar\Gamma^\ml}}

\def\Gpmb {{\Gamma^\pmb}}
\def\Gpnb {{\Gamma^\pnb}}
\def\Gppb {{\Gamma^{+\bar p}}}
\def\Gmm {{\Gamma^\mm}}
\def\Gmn {{\Gamma^{-n}}}
\def\Gmp {{\Gamma^{-p}}}

\def\bp {{\beta^+}}
\def\bm {{\beta^-}}
\def\gp {{\gamma^+}}
\def\gm {{\gamma^-}}
\def\tpa {{\theta_+^{\alpha}}}
\def\tma {{\theta_-^{\alpha}}}
\def\tbpa {{\bar\theta_+^{\alpha}}}
\def\tbma {{\bar\theta_-^{\alpha}}}

\def\tmb {{\theta_-^{\beta}}}

\tolerance=5000
\footline={\ifnum\pageno>1
       \hfil {\rm \folio} \hfil
    \else \hfil \fi}

\overfullrule=0pt 
\baselineskip=18pt
\raggedbottom
\centerline{\bf Finiteness and Unitarity of}
\centerline{\bf Lorentz-Covariant Green-Schwarz Superstring Amplitudes}
\vskip 24pt
\centerline{Nathan Berkovits}
\vskip 24pt
\centerline{Math Dept., King's College, Strand, London, WC2R 2LS, United
Kingdom}
\vskip 12pt
\centerline{e-mail: udah101@oak.cc.kcl.ac.uk}
\vskip 12pt
\centerline {KCL-TH-93-6}
\vskip 12pt
\centerline {March 1993}
\vskip 96pt
\centerline {\bf Abstract}
\vskip 12pt

In two recent papers, a new method was developed for calculating
ten-dimensional superstring amplitudes with an arbitrary number
of loops and external massless particles, and for expressing them in
manifestly Lorentz-invariant form. By explicitly checking for
divergences when the Riemann surface degenerates, these amplitudes
are proven to be finite. By choosing light-cone moduli for the surface
and comparing with the light-cone Green-Schwarz formalism,
these amplitudes are proven to be unitary.
\vfil
\eject
\centerline{\bf I. Introduction}
\vskip 12pt

There are various methods for calculating ten-dimensional
superstring amplitudes, each with its
advantages and disadvantages. One method is
the light-cone Neveu-Schwarz-Ramond formalism,$^{1}$ which has the advantage
of being manifestly unitary,
but the disadvantage of non-trivial
operators at the interaction points of the light-cone diagram.
Besides complicating considerably the amplitude calculations, the
presence of these non-trivial interaction-point operators forces
the introduction of higher-order contact terms,$^{2,3}$ which are
necessary to remove the unphysical divergences when two or more
interaction points coincide.

Another method for calculating superstring
amplitudes is the manifestly Lorentz-covariant Neveu-Schwarz-Ramond
formalism.$^4$ In place of the interaction-point operators of the
light-cone formalism, the covariant Neveu-Schwarz-Ramond formalism
contains N=1 fermionic moduli, which after integrating out gives
rise to picture-changing operators. These picture-changing operators
differ from the light-cone interaction-point operators in that their
locations on the surface are not predetermined.

Moving the locations of the picture-changing operators changes the
integrand of the scattering amplitude by a total derivative,$^5$ which
gives a surface term contribution due to the presence of a cutoff
in the bosonic moduli (the cutoff is necessary in the Neveu-Schwarz-Ramond
formalism since before summing over spin structures, the integrand is
divergent at points in moduli space where the surface degenerates).$^6$
Although this surface-term ambiguity can be removed by requiring that
that the scattering amplitudes agree with those obtained using the
manifestly unitary light-cone Neveu-Schwarz-Ramond formalism (this implies
that at divergent points in moduli space, the locations of the
picture-changing operators should coincide with the interaction points
of the corresponding light-cone diagram), the need to impose such a
requirement is an unpleasant feature of the covariant Neveu-Schwarz-Ramond
formalism. Note that because the light-cone interaction-point
locations are not holomorphic functions of the moduli, this requirement
can not be imposed independently on the right and left-moving contributions
to the amplitude.

It is also possible to calculate superstring amplitudes without
integrating out the fermionic moduli, using either the light-cone$^{7,8}$
or covariant$^9$ Neveu-Schwarz-Ramond formalism on supersheets. With either
of these two methods, one ends up with a Lorentz-covariant super-integrand
which needs to be integrated over a super-moduli space with boundary.
The shape of this boundary can be determined by requiring that the
bosonic moduli of the corresponding light-cone super-diagram are
pure complex numbers with no nilpotent parts, which guarantees that
the light-cone supersheet and non-supersheet formalisms give equivalent
amplitudes.$^7$

Another approach to calculating superstring amplitudes is to use the
Green-Schwarz formalism, which is manifestly spacetime-supersymmetric
and requires no sum over spin structures. Until recently, the only
method available for caculating Green-Schwarz superstring amplitudes
was the light-cone method,$^{10,11,12}$
which is manifestly unitary but requires
non-trivial operators at the interaction points of the light-cone
diagram (note that the semi-light-cone method$^{13}$ requires
the same non-trivial operators when $\dz x^{9+0}=0$, so
it contains no advantages over the light-cone method$^{14}$).
Because of complications caused by these interaction-point
operators, the only manifestly Lorentz-invariant scattering amplitudes
that have been calculated using the light-cone Green-Schwarz method
have been the tree and one-loop amplitude with four external
massless particles. Furthermore, proofs of finiteness and Lorentz
invariance using the light-cone Green-Schwarz method are complicated
by the presence of non-physical singularities when two or more
interaction-point operators collide.$^{2,3,12}$ Although these non-physical
singularities can be removed by introducing higher-order contact
terms, it has not yet been proven that the resulting scattering
amplitudes are Lorentz invariant.\footnote
\dag{ Restuccia and Taylor have claimed that the heterotic superstring
amplitudes are finite and Lorentz-invariant after introducing only
fourth-order contact terms,$^{12}$ however their claim is based on mistakenly
analyzing the non-physical singularities as a function of the distance
between nearby interaction points, $\nu$, rather than as a function of
the Lorentz-invariant
positions of the punctures, $z_r$. Since $\nu$ is proportional to
$\sqrt {z_r -y}$ for some $y$ as $\nu \to 0$,$^{15}$ the behavior
$\nu^{-3} \bar \nu d^2\nu$ of equation (4.60a) in reference 12
becomes $(z_r -y)^{-2} d^2 z_r$, which has the expected holomorphic divergence
(when interaction points collide,
the right-moving sector of the heterotic superstring has the same
regular behavior as the bosonic string). Note that the manifestly
Lorentz-invariant amplitudes in equation (3.52) of reference 12
require contact-term contributions similar to those mentioned in
Section 3.10 of reference 12.}

An indirect way of proving Lorentz invariance of the light-cone
Green-Schwarz formalism is to prove equivalence with the light-cone
Neveu-Schwarz-Ramond formalism (after summing over spin structures),
which has already been proven equivalent to the manifestly covariant
Neveu-Schwarz-Ramond formalism$^9$ (after imposing the previously mentioned
non-holomorphic condition near divergent points in moduli space). This
indirect link also proves finiteness of the covariant
Neveu-Schwarz-Ramond formalism, since the only dangerous divergence in
the covariant Neveu-Schwarz-Ramond amplitudes comes from the dilaton
tadpole,$^7$ which is easily seen to vanish in the light-cone
Green-Schwarz formalism (after assuming Lorentz invariance).$^{12,16}$
Although it is not difficult to prove the equivalence of the
light-cone Green-Schwarz and light-cone
Neveu-Schwarz-Ramond formalisms (the three-string vertex is
the same in the two formalisms,$^{11}$ and
summing over spin structures in the light-cone Neveu-Schwarz-Ramond
formalism correctly performs the GSO projection$^3$),
it would be preferable to have a more
direct proof that Lorentz-covariant superstring amplitudes are finite.

Recently, a new method has been developed for calculating Green-Schwarz
superstring amplitudes that does not require light-cone gauge fixing,
and therefore does not contain the problems of light-cone interaction-point
operators.$^{17}$
Although this formalism is not manifestly Lorentz invariant,
it is straightforward to construct covariant vertex
operators and the full set of SO(9,1) Lorentz
generators, and to write manifestly Lorentz-invariant expressions
for scattering amplitudes involving an arbitrary number of loops
and external massless particles.$^{18}$

In this paper, these manifestly Lorentz-invariant Green-Schwarz amplitudes
(for external massless bosons)
will be proven to be finite by explicitly checking that all possible
divergences are absent. By parameterizing the Riemann surface with
light-cone moduli, it will then be proven that these superstring
amplitudes agree with amplitudes obtained using the light-cone
Green-Schwarz formalism. This proves that the finite Lorentz-invariant
Green-Schwarz scattering amplitudes are unitary and that the
light-cone Green-Schwarz amplitudes are
Lorentz invariant.
Note that this proof of equivalence with the light-cone formalism
will not rely on any assumptions such as those made in reference 17.

Although only the Type IIB Green-Schwarz superstring will be discussed
in this paper, it should be straightforward to generalize these results
to all other types of Green-Schwarz superstrings.

\vskip 12pt
\centerline{\bf II. Lorentz Covariant Green-Schwarz Superstring Amplitudes}
\vskip 12 pt

This section will summarize the results of references 17 and 18 in which
Lorentz-covariant Green-Schwarz superstring amplitudes were calculated.

The free matter fields on a
Euclidean worldsheet needed to describe the Type IIB Green-Schwarz
superstring consist of ten real spin 0 bosons, $x^\mu$ ($\mu=0$ to 9),
four pairs of right-moving spin $\half$ fermions, $\Gplb$ and $\Gml$
($l=1$ to 4),
two pairs of right-moving spin 0 and spin 1 fermions,
$\psi^\pm$ and $\varepsilon^\mp$,
one pair of right-moving spin 0
bosons, $h^+$ and $h^-$; and their complex conjugates,
four pairs of left-moving spin $\half$ fermions, $\Gbplb$ and $\Gbml$,
two pairs of left-moving spin 0 and spin 1 fermions,
$\bar\psi^\pm$ and $\bar\varepsilon^\mp$,
and one pair of left-moving spin 0
bosons, $\bar h^+$ and $\bar h^-$.
The chiral bosons, $h^\pm$ and $\bar h^\pm$, all have screening charge
$-1$ and take values on a circle of radius 1.

The action for these free matter fields contains an N=(2,2) superconformal
invariance$^{19}$ which can be used to gauge-fix $x^0$, $x^9$, and all
$\psi$'s, $\varepsilon$'s, and $h$'s, leaving only the usual light-cone
Green-Schwarz fields of eight $x$'s, eight $\Gamma$'s, and eight
$\bar \Gamma$'s. The generators for the right-moving N=2
superconformal transformations are:

$$T=\Gplb\Gml -\dz\hp +\dz\hm,\quad
G_-=\dz\xplb\Gml+(\vep +\half\psi^+\dz x^{9-0}) e^{-\hp},\quad
G_+=\dz\xml\Gplb+ \eqno(II.1)$$
$$(\vem+\half \dz x^{9-0}\psi^- )
((\dzxp)
 e^\hp +e^{-\hm})-e^\hp ((\dz h^+ +\dz h^-)\dz\psi^- +{3\over 4}
\partial^2_z \psi^-),$$
$$L=
\dz x^\plb \dz x^\ml -\half (\Gplb\dz\Gml +\Gml\dz\Gplb)-\vep
\dz\psi^- -\vem\dz\psi^+ +\dz \hp\dz\hm +\half(\partial^2_z \hp
+\partial^2_z \hm ),$$
where $x^\plb\equiv x^l +i x^{l+4}$,
$x^\ml\equiv x^l -ix^{l+4}$, and $x^{9\pm 0}\equiv x^9 \pm x^0$.

The N=(2,2) ghost and anti-ghost fields coming from gauge-fixing these
invariances consist of a pair of right-moving fermions of spin $-1$
and $+2$, $c$ and $b$, two pairs of right-moving bosons of spin $-\half$
and $+{3\over 2}$, $\gamma^\pm$ and $\beta^\mp$, a pair of
right-moving fermions of spin 0 and 1, $u$ and $v$; and their left-moving
complex conjugates, $\bar c$, $\bar b$, $\bar\gamma^\pm$, $\bar\beta^\mp$,
$\bar u$, and $\bar v$. Since the central charge contribution of the
matter fields cancels the contribution of the ghost fields, a nilpotent
BRST charge $Q$ can be constructed in the usual way out of the N=2
stress-energy tensor and the ghosts.$^{20}$

It is convenient to bosonize the bosonic ghosts,$^4$
$$\gp=e^{\php}\etp,\quad \bm= e^{-\php}\dz\xm,
\quad\gm=e^{\phm}\etm,\quad \bp= e^{-\phm}\dz\xp,\eqno(II.2)$$
where $\eta^\pm$ and $\xi^\mp$ are a pair of spin 1 and spin 0
fermions, and $\phi^\pm$ are two scalar bosons of screening charge $+2$
with negative energy (i.e.,
$\dy\php (y) \dz\php (z)$
$\to -(y-z)^{-2}$).

Picture-changing operators, $F^\pm$, can then be defined in the following
way:
$$F^+\equiv[Q,\xp]=e^\phm [G^+_{matter} +
(b-\half\dz v)\gamma^+ -v\dz\gp]+c\dz\xp ,
\eqno(II.3)$$
$$F^-\equiv[Q,\xm]=e^\php [G^-_{matter} +
(b+\half\dz v)\gamma^- +v\dz\gm]+ c\dz\xm ,$$
where $G^\pm_{matter}$ is defined in equation (II.1).

Instanton-number changing operators, $I$ and $I^{-1}$, can also
be defined as
$$I=e^{\int [Q,v]} =\epsilon^{lmnp}\Gplb\Gpmb\Gpnb\Gppb
e^{h^- -h^+ +\phm -\php +cv},\eqno(II.4)$$
$$I^{-1}=e^{-\int [Q,v]} =\epsilon_{lmnp}\Gml\Gmm\Gmn\Gmp
e^{h^+ -h^- +\php -\phm -cv}.$$
Just as the ghost-number of an operator is defined by commuting with
the ghost charge, $\int dz(cb +uv+\dz\php +\dz\phm)$, the instanton-number
of an operator is defined by commuting with the instanton charge,
$\half \int dz (\vep \sm-\vem\sp +\dz h^- -\dz h^+)$. It is easily
checked that $F^\pm$ and $I^{\pm 1}$ are in the BRST cohomology,
but $\dz F^\pm$ and $\dz I^{\pm 1}$ are BRST trivial.

In order to prove Lorentz invariance,
one needs to construct a generalization of the SO(9,1) generators,
$m^{\mu\nu}=\int dz (x^\mu \dz x^\nu)-\int d\bar z (x^\mu \dzbar x^\nu)$,
which is BRST invariant. This generalization is
$$M^{\mu\nu}=
\int dz \{ G^+ [G^- ,x_+^\mu x_-^\nu ]\}
-\int d\bar z \{\bar G^+ [\bar G^- ,\bar x_+^\mu \bar x_-^\nu ]\},
\eqno(II.5)$$
$$where \quad
x_-^{9+0}=x^{9+0} -\half \sp\sm,\quad
x_-^{9-0}=x^{9-0} +e^{\hp +\hm},\quad
x_-^\plb =x^\plb -e^\hm \sp\Gplb,\quad
x_-^\ml =x^\ml,$$
$$x_+^{9+0}=x^{9+0} +\half \sp\sm,\quad
x_+^{9-0}=x^{9-0},\quad
x_+^\plb =x^\plb,\quad
x_+^\ml =x^\ml-e^\hp\sm\Gml,$$
and
$G^\pm$ is defined in equation (II.1) (note that $\{ G^+, x_-^\mu\}=
\{ G^-, x_+^\mu\}=0$). It is straightforward to check that
$M^{\mu\nu}$ generates an SO(9,1) algebra, that $x_{\pm\pm}^\mu\equiv
x_{\pm}^\mu+\bar x_{\pm}^\mu -x^\mu$
and $x_{\pm\mp}^\mu\equiv
x_{\pm}^\mu+\bar x_{\mp}^\mu -x^\mu$ transform as SO(9,1) vectors,
and that $\sp$, $\sm$, $\sbp$, $\sbm$ transform as one component
of sixteen-component SO(9,1) Weyl spinors, $\tpa$, $\tma$, $\tbpa$,
$\tbma$ (see reference 18 for their explicit form).

The covariant vertex operators with ghost-number ($-1,-1$) and
instanton-number ($m,\bar m$) for the massless particles are:
$$V_{m\bar m} (z,\bar z) = p_{AB}\quad
 c \bar c\quad e^{-(\phi^+ +\phi^- +
\bar\phi^+ +\bar\phi^-)}\quad
 V_m^A \bar V_{\bar m}^B\quad  e^{-i k_\mu x^\mu},\eqno(II.6)$$
where $A$ and $B$ are either SO(9,1) vector-indices $\mu$
or sixteen-component
SO(9,1) Weyl spinors $\alpha$, $p_{AB}$ is the polarization tensor satisfying
$k^\mu p_{\mu B}=k^\mu p_{A \mu}=0$, and
$V_0^\mu=\tpa \gmu \tmb
e^{ik_\nu x^\nu}$, $V_{\pm\half}^\alpha =\theta_\pm^\alpha e^{ik_\mu
x_\mp^\mu}.$

On a genus $g$ surface with period matrix
$\tau$ and zero instanton number, the scattering
amplitude of N external massless states is:
$$A=
|\prod_{i=1}^{3g-3+N}\int dm^T_i
\prod_{j=1}^g \int dm^{U(1)}_j |^2 \eqno(II.7)$$
$$\int (\prod_{\mu=0}^9 Dx^\mu ) |(\prod_{l=1}^4 D\Gplb D\Gml )
D\sp D\vem D\sm D\vep D\hp D\hm Dc Db D\gamma^+ D\gamma^- D\beta^+
D\beta^- |^2 $$
$$|\prod_{i=1}^{3g-3+N}\int d^2 y_i (M^i (y_i) b(y_i))
(\prod_{k=1}^{2g-2+N} F^+ (w_k^+) F^- (w_k^-))  Z_1 (\tau) |^2
e^{-S}\prod_{r=1}^N
V_{m_r,\bar m_r} (z_r,\bar z_r) ,$$
where $m^T_i$ and $M^i$ are the usual bosonic Teichmuller parameters
and their Beltrami differentials, $m^{U(1)}_j$ are the $g$ U(1)
moduli which range over the Jacobian variety $C^g /(Z^g +\tau Z^g)$
and measure the change in phase $e^{irm^{U(1)}_j }$ when a field
of U(1) charge $r$ goes around the $j^{th}$ $b$-cycle of the surface,
$F^\pm$ are the picture-changing operators defined in equation (II.3)
which come from integrating over the fermionic moduli, $w_k^\pm$
are arbitrary points on the surface chosen independently of the
bosonic moduli, the term $Z_1 (\tau)$ comes from performing the
path integral over the $(u,v)$ ghosts (this path integral is
trivial after inserting a zero-mode for $u$, since no $u$ fields
appear in $F^\pm$ or $V$), the vertex operators are chosen such
that $\sum_{r=1}^n m_r
=\sum_{r=1}^n \bar m_r=0$, and $S$ is the Wick-rotated
free-field action.

The path integrals over the fermions can be
performed by bosonizing and using the formula:$^{21}$
$$\int D(e^\phi )D(e^{-\phi} ) e^{-S(\phi)}\prod_{i=1}^n \exp
(c_i \phi(z_i))=Z(\sum_{i=1}^n c_i z_i ;q,r;\tau) \eqno(II.8)$$
$$=\delta_{q(g-1),\Sigma c_i}
\prod_{i<j} E(z_i, z_j)^{c_i c_j} \prod_{i=1}^n
\sigma (z_i)^{q c_i}
(Z_1(\tau))^{-\half}
\Theta([\sum_{i=1}^n c_i z_i -q\Delta]-r m^{U(1)},\tau),$$
where $q$ and $r$ are the screening charge and
the U(1) charge for $e^\phi$, $\Delta$ is the Riemann class,
$\tau$ is the period matrix of the surface and is a complex
symmetric $g\times g$ matrix with positive-definite
imaginary part,
$Z_1(\tau)$ is a normalization for $Z$ such that
$Z(\sum_{i=1}^g z_i -y;1,0;\tau)=Z_1(\tau)
\det_{ij} \omega_i (z_j)$, $\omega_i$ are the g canonical holomorphic
one-forms satisfying $\int_{a_j}\omega_i=\delta_{i,j}$ and
$\int_{b_j}\omega_i=\tau_{ij}$, $E(y,z)$ is the prime form,
${{\sigma (y)}\over {\sigma (z)}}=
{{\Theta([y-\sum_{i=1}^g p_i +\Delta],\tau)}\over
{\Theta([z-\sum_{i=1}^g p_i +\Delta],\tau)}}
\prod_{j=1}^g {{E(y,p_j)}\over{E(z,p_j)}}$
for arbitrary $p_i$
(the final amplitudes will contain equal powers of $\sigma$ in the
numerator and denominator because of the vanishing conformal anomaly),
$[\sum_{i=1}^n (y_i -z_i)]_j\equiv\sum_{i=1}^n \int_{z_i}^{y_i}
\omega_j$ is an element in the Jacobian variety $C^g/ (Z^g+\tau Z^g)$,
and
$$\Theta(z,\tau)\equiv
\sum_{n\in Z^g} \exp(i\pi n_j\tau_{jk} n_k+2\pi in_j z_j) \eqno(II.9)$$
which satisfies
$\Theta(z+\tau n+m,\tau)=
\exp(-i\pi n_j\tau_{jk} n_k-2\pi in_j
z_j)
\Theta(z,\tau)$ for $z\in C^g$ and $n,m\in Z^g$.
For a brief but sufficient review of these objects, see Chapter 3
of reference 21.

The formulas for the path integrals over the bosonic fields are:$^5$
$$\int Dx^\mu e^{-S(x^\mu)}
\prod_{j=1}^n \exp(ip_j x^\mu (z_j))
= \delta (\sum_{j=1}^n p_j)
(\det Im\, \tau)^{-\half} |Z_1(\tau)|^{-1}
\prod_{j<k} F(z_j,z_k)^{p_j p_k}
,\eqno(II.10)$$
where $F(y,z)=\exp(-2\pi Im[y-z](Im\, \tau)^{-1} Im[y-z]) |E(y,z)|^2$,
$$\int D\beta^+ D\gamma^- e^{-S(\beta^+,\gamma^-)} \prod_{i=1}^m
\xp (x_i)\prod_{j=1}^n \eta^- (y_j)\prod_{k=1}^p \exp (c_k \phi^-(z_k))
=\eqno(II.11)$$
$${\prod_{l=1}^n Z(-y_l+\sum_{i=0}^m x_i -\sum_{j=1}^n y_j +\sum_{k=1}^p
c_k z_k\quad;2,-1;\tau)}\over
{\prod_{l=0}^m Z(-x_l+\sum_{i=0}^m x_i -\sum_{j=1}^n y_j +\sum_{k=1}^p
c_k z_k\quad;2,-1;\tau)},$$
where the location of the $\xp$ zero-mode, $x_0$, can be chosen anywhere
on the surface,$^{22}$ and
$$\int D\hp D\hm e^{-S(\hp,\hm)}\prod_{i=1}^m \exp
(c^-_i \hp (y_i))\prod_{j=1}^n (c^+_j \hm (z_j))=\eqno(II.12)$$
$$(Z_1(\tau))^{-1}\delta_{1-g,\Sigma c^-_i}
\delta_{1-g,\Sigma c^{+}_i}
\prod_{i\not= j} E(z_i,z_j)^{c^{-}_i c^{+}_j}
\prod_{i=1}^m \sigma(z_i)^{-c_i^-}
\prod_{j=1}^n \sigma(z_j)^{-c_j^+}
\prod_{k=1}^g \delta (m^{U(1)}_k -
[\sum_{j=1}^n c^{+}_j z_j +\Delta]_k),$$
where the constraint imposed on the U(1) moduli can be understood as
coming from the global constraint, $\Omega$, of equation (II.4) in
reference 17.
Note that both $(\beta^\pm,\gamma^\mp)$
and $h^\pm$ contain fields with negative energy (i.e.,
$\dy\phi^\pm (y) \dz\phi^\pm (z)$ and $\half (\dy\hp(y) -\dy\hm(y))
(\dz\hp(z)-\dz\hm(z))$ go like $-(y-z)^2$ as $y \to z$), which
can cause unphysical poles in the path integrals. For the $(\beta^\pm,
\gamma^\mp)$ path integral, the residues of these poles are BRST
trivial;$^5$ for the $(\hp,\hm)$ path integral, these unphysical poles
are avoided by the $\delta$-function restriction on the U(1) moduli.

In order to make the Lorentz invariance of the amplitude manifest
as a function of the momenta $k_r^\mu$ ($r=1$ to N) and the polarizations
$p^r_{AB}$, it is convenient to introduce a null real SO(9,1) vector,
$m^\mu$, and a pure complex SO(9,1) spinor, $v^\alpha$, satisfying
$\va\gmu\vb =\vba\gmu\vbb=\m m_\mu=0\quad\hbox{  and  }\quad
\va\gmu\vbb m_\mu=1.$
By using this vector and spinor to break SO(9,1) down to SU(4)
(e.g. $k^{9+0}=(v\gamma_\mu \bar v) k_r^\mu$), the amplitude $A$ can be
written as a manifestly Lorentz-covariant function of $m^\mu$,
$v^\alpha$, $\bar v^\alpha$, $k_r^\mu$, and $p^r_{AB}$, where $A$
is a polynomial in $m^\mu$, $v^\alpha$, and $\bar v^\alpha$
(note that
$m^\mu$, $v^\alpha$, and $\bar v^\alpha$ appear only in the nilpotent
part of $k_\mu x^\mu_\pm$ since the non-nilpotent part of $x^\mu_\pm$
is simply $x^\mu$).

Since $A$ is guaranteed to also be a Lorentz-covariant function of
just the
$k_r^\mu$'s and $p^r_{AB}$'s ($M^{\mu\nu}$ is BRST invariant and the
vertex operators transform covariantly), all monomials of
$m^\mu$, $v^\alpha$, and $\bar v^\alpha$ can be replaced by
their Lorentz-invariant component. For example,
$m^\mu v^\alpha\bar v^\beta\to {1\over 160}
(\gamma^{\mu\,\alpha\beta})$, and
$m^\mu m^\nu v^\alpha v^\beta
\bar v^\gamma\bar v^\delta\to$
$ {1\over 31300} (\gamma^{\mu\,\alpha\gamma}
\gamma^{\nu\,\beta\delta}+
\gamma^{\mu\,\alpha\delta}
\gamma^{\nu\,\beta\gamma}+
\gamma^{\mu\,\beta\gamma}
\gamma^{\nu\,\alpha\delta}+
\gamma^{\mu\,\beta\delta}
\gamma^{\nu\,\alpha\gamma}+$
$ {1\over 4}(
\gamma^{\rho\,\alpha\beta}
\gamma^{\gamma\delta}_\rho \eta^{\mu\nu}-
\gamma^{\mu\,\alpha\beta}
\gamma^{\nu\,\gamma\delta}-
\gamma^{\mu\,\gamma\delta}
\gamma^{\nu\,\alpha\beta})).$
In this way, the amplitude $A$ can be written as a manifestly
Lorentz-invariant function of just the
$k_r^\mu$'s and $p^r_{AB}$'s.$^{18}$

\vfil
\eject
\centerline{\bf III. Finiteness}
\vskip 12 pt

In order to check for possible divergences in $A$ of equation (II.7),
it is convenient to choose the ``tree-with-tadpoles'' parameterization
for the genus $g$ Riemann surface with N external states.$^{23}$ This
parameterization consists of sewing together $2g-2+N$ spheres, $S_i$,
each having three punctures, $P_{i,r}$ ($i=1$ to $2g-2+N$, $r=1$ to 3).

First sew $P_{i,3}$ to $P_{i+1,1}$ for $i=1$ to $g-1$ where the
radius of the sewed puncture is $B_i$ ($B_i$ is a complex number whose
phase rotates one puncture with respect to the other). Next sew
$P_{i+g-1,3}$ to $P_{i+g,1}$ for $i=1$ to $N-1$ where the radius
of the sewed puncture is $L_i$. Now sew $P_{i+g+N,3}$ to $P_{i+1,2}$
for $i=1$ to $g-2$ where the radius of the puncture is $H_i$, and
glue $P_{g+N-1,3}$ to $P_{g+N,1}$ where the radius of the puncture is
$H_{g-1}$. At this point, spheres $S_1 ... S_{g+N}$ are sewn together,
and sphere $S_{i+g+N}$ is sewn to sphere $S_{i+1}$ for $i=1$ to $g-2$.
Finally, sew $P_{i+g+N,1}$ to $P_{i+g+N,2}$ with radius $K_i$ for
$i=1$ to $g-2$,
sew $P_{g+N,2}$ to $P_{g+N,3}$ with radius $K_{g-1}$,
sew $P_{1,1}$ to $P_{1,2}$ with radius $K_{g}$, and insert the
$N$ external states on the remaining unsewed punctures $P_{g,2} ...
P_{g+N-1,2}$.

The $3g-3+N$ complex bosonic Teichmuller parameters for this
parameterization are given by the complex radii, $B_i$ ($i=1$ to $g-1$),
$L_i$ ($i=1$ to $N-1$),
$H_i$ ($i=1$ to $g-1$), and
$K_i$ ($i=1$ to $g$).
For this choice of moduli, the contribution from the Beltrami differentials
is$^{23}$
$$| \prod_{i=1}^{3g-3+N} \int_{C_i} {y_i b(y_i)\over R_i} dy_i |^2,
\eqno(III.1)$$
where $C_i$ is a closed loop surrounding the sewed punctures that
have the radius $R_i$, and $y_i=0$ at the
center of these sewed punctures.
The locations of the picture-changing operators will be chosen such
that one $F^+, F^-, \bar F^+,$ and $\bar F^-$ sits on each of the
$2g-2+N$ spheres, but are otherwise arbitrary. Choose the extra
zero mode of the $\xi^\pm$ and $\bar\xi^\pm$ fields to sit on $S_{g+N-1}$.

Since the path integral over the $h^\pm$ fields determines the values
of the U(1) moduli $m^{U(1)}_j$, the only possible divergence in $A$
can arise at limiting points of the radii $B_i$, $L_i$, $H_i$, and
$K_i$.\footnote\dag{
The non-physical poles coming from the zeroes of the $\Theta$-functions
in equation (II.11) have residues which are total derivatives in
the Teichmuller parameters (this is clear since by shifting the
picture-changing operators by a BRST trivial quantity, $\{Q,\xi^\pm\}$,
which changes the
integrand of $A$ by a total derivative, the locations of these non-physical
poles can be altered).$^5$ So if the amplitude is finite near the limiting
points of the radii, there is no need to introduce a cutoff for the
Teichmuller parameters, the moduli space has no boundary, and these
poles are harmless.}
Because of modular invariance (as was shown in Section IV.A. of
reference 17, the integrand of $A$ is independent of the spin-structure
chosen for the fields of $\half$-integer conformal weight), one
only needs to check for divergences when the radii approach zero.$^{24}$

When $K_i\to 0$, which corresponds to the $i^{th}$ $a$-cycle on the
surface being shrunk to zero, $\tau_{ii}\to {1\over 2\pi i} \log K_i$ and
$\Delta_i\to -\half \tau_{ii}$.$^{25}$ From
the definition of the $\Theta$-function
in equation (III.9) and
from the fact that $m^{U(1)}_i=
[\sum_{j=1}^n c^{+}_j z_j +\Delta]_i)$, one finds that $Z(q,r;\tau)$
diverges like $(K_i^{-\half(q+r-1)}+1)$ and that $\det Im \tau$ diverges
like $\log |K_i|$. So the path integral over $(b,c)$ diverges like
$K_i^{-1}$, the path integral over $(\beta^-,\gamma^+)$ converges
like $K_i$, the ten path integrals over $x^\mu$ each converge like
$(\log |K_i|)^{-\half}$,
and all other path integrals are regular. After combining
with the $|K_i|^{-2}$ dependence from the Beltrami differentials, one
finds that
$$A \to (\log |K_i|)^{-5} |K_i|^{-2} d^2 K_i ,\eqno(III.2)$$
which is not divergent as $K_i\to 0$ (for $y=(log |K_i|)^{-1}$,
$A \to y^3 dy$ as $y\to 0$).

When any other radius $R$ is shrunk to zero, the genus $g$ surface
with $N$ punctures degenerates into a genus $g_1$ surface with $N_1$
punctures and a genus $g_2$ surface with $N_2$ punctures. These
two surfaces, $G_1$ and $G_2$, have period-matrices, $\tau_1$ and
$\tau_2$, such that the original period matrix, $\tau$, decomposes
into the direct-sum of $\tau_1$ and $\tau_2$.
As was shown in reference 21, this implies that
$$E(x,y)\to  R^{-\half} E_1 (x, p_1) E_2 (y, p_2),
\quad
\sigma (x)\to R^{\half g_2 a_2} \sigma_1(x)
\sigma_1(p_1)^{a_1 -1}
\sigma_2(p_2)^{a_2} E_1(x,p_1)^{-g_2},\eqno(III.3) $$
$$Z(\sum_{i=1}^m  c_i x_i +
\sum_{j=1}^n  d_j y_j ;q,r;\tau)\to R^{-\half q_1 q_2}
Z(\sum_{i=1}^m  c_i x_i -q_1 p_1;q,r;\tau_1)
Z(\sum_{j=1}^n  d_j y_j -q_2 p_2;q,r;\tau_2),$$
where $x$ is on $G_1$, $y$ is on $G_2$, $p_i$ is the location of the
sewed puncture with radius $R$ on $G_i$, $a_i={g_i -1 \over g-1}$,
$q_1=\sum_{i=1}^m c_i-q(g_1-1)$, $q_2=
\sum_{j=1}^n d_j-q(g_2-1)$, and $q_1 +q_2=q$.

For a given $R$, choose $G_2$ to be the surface containing
$S_{g+N-1}$, and push the loop containing the Beltrami
differential for $R$ onto $G_2$. Since there are $3g_1-2+N_1$
$b$ fields coming from Beltrami differentials on $G_1$ and
$N_1$ $c$ fields coming from vertex operators on $G_1$,
the $(b,c)$ path integral behaves as $R^{\half n_b (n_b -3)}$,
where $n_b -1$ is the number of $b$'s minus the number of $c$'s
coming from the $2(2g_1-1+N)$ picture-changing operators $F^\pm$
on $G_1$. Similarly, the $(\beta^\pm, \gamma^\mp)$
path integrals each behave as $R^{\half (1-n_{\xi^\pm})}$, where
$n_{\xi^\pm}$ is the number of $\xi^\pm$'s minus the number
of $\eta^\mp$'s coming from $G_1$. Since $n_\xp +n_\xm =1-n_b$,
the combined path integral behaves as $R^{\half(n_b -1)^2}.$

The path integral over the $(\hp,\hm)$ fields behaves
like $R^{{1\over 4}[(n_\hp +n_\hm +2)(n_\hp +n_\hm)-(n_\hp -n_\hm)^2]}$,
and the path integral over the $(\psi^\pm,\varepsilon^\mp)$ behaves
like $R^{{1\over 4}[(n_\sp +n_\sm +2)(n_\sp +n_\sm)+(n_\sp -n_\sm)^2]}$,
where $(n_{h^\pm} +1-g)$ is the number of $e^{h^\pm}$ terms on $G_1$
and $(n_{\psi^\pm}+1-g)$ is the number of $\psi^\pm$'s minus the
the number of $\varepsilon^\mp$'s on $G_1$.
Since all operators on the surface have zero instanton charge
(it has been assumed that all vertex operators are boson-boson
vertex operators of instanton-number (0,0)), $n_\sp- n_\sm =
n_\hm -n_\hp$, so the combined path integral behaves like
$$R^{{1\over 4}[(n_\hp +n_\hm +2)(n_\hp +n_\hm)
+(n_\sp +n_\sm +2)(n_\sp +n_\sm)]}.\eqno(III.4)$$

Finally, the path integrals over the $(\Gml,\Gplb)$ fields behave like
$R^{\half n_l^2}$ where $n_{l}$ is the number of $\Gml$ fields
minus $\Gplb$ fields on $G_1$, and the path integrals over the
$x^\mu$ fields behave like $|R|^{k^\mu k_\mu}$ where $k^\mu$
is the momentum crossing from $G_1$ to $G_2$.

The only way that the expression in equation (III.4) can diverge is
if $n_\hp+n_\hm =n_\sp+n_\sm=-1$, in which case it behaves like
$R^{-\half}$ (note that because of zero instanton charge,
$n_\hp+n_\hm$ must be odd if $n_\sp+n_\sm$ is odd). However, if
$n_\hp +n_\hm$ is odd, then $n_\hp-n_\hm$ must also be odd,
implying by U(1) conservation that either $n_\xp -n_\xm$ is odd
(which implies that $n_b$ is not equal to one) or at least one
of the $n_l$'s is non-zero. In either case, the other path
integrals behave at worst like $R^{+\half}$.

So after combining these results with the $d^2 R/|R|^2$ dependence of the
Beltrami differentials, one finds that as $R\to 0$,
$$A \to A_1 A_2  |R|^{k^\mu k_\mu -2} \quad d^2 R ,\eqno(III.5)$$
where $k^\mu=
\sum_{r=1}^{N_1} k_r^\mu=-
\sum_{r=N_1+1}^{N} k_r^\mu$, and
$A_1$ is the amplitude on $G_1$ with $4(2g-1+N_1)$ picture-changing
operators, $3g-2+N_1$ loops from
Beltrami differentials, $N_1$ vertex operators,
and an operator sitting at the
location of the shrunk puncture $p_1$ of the form:
$$|c^{n_b} e^{-\phi^+ -\phi^-}(\gamma^-)^{n_\xp}
(\gamma^+)^{n_\xm} (\vem)^{n_\sp}
(\vep)^{n_\sm}e^{-n_\hp \hp} e^{-n_\hm \hm}
\prod_{l=1}^4 (\Gplb)^{n_l}|^2 e^{ik_\mu x^\mu} .\eqno(III.6)$$

If $R$ is $L_i$, then $N_i =i$, $g_1=g-1$, and the scattering
amplitude has the expected massless pole when $k_\mu k^\mu=0$.
If $R$ is not one of the $L_i$'s, then $N_1=k^\mu=0$,
and it will now be shown that this implies $A_1$ is zero.

After doing the field redefinition,
$$[x^{9+0}\to x^{9+0}+\half\sp\sm,
\vep\to \vep+\half\sp\dz x^{9-0},
\vem\to \vem-\half\sm\dz x^{9-0}],\eqno(III.7)$$
the picture-changing operators, $F^\pm$, no longer contain
the zero mode of $\sp$ (note that this redefinition process
does not affect the measure factor since it preserves the
free-field commutation relations). Similarly,
after doing the field redefinition,
$$[x^{9+0}\to x^{9+0}-\half\sp\sm,
\vep\to \vep-\half\sp\dz x^{9-0},
\vem\to \vem+\half\sm\dz x^{9-0}],\eqno(III.8)$$
the picture-changing operators, $F^\pm$, no longer contain
the zero mode of $\sm$.
Since $A_1$ is zero unless there is a zero mode for $\sp$ and
$\sm$ somewhere on the surface, $n_\sp$ and $n_\sm$ must both
be less than zero (these zero modes of $\psi^\pm$ are related to
spacetime supersymmetry since two of the sixteen right-moving
spacetime-supersymmetry generators are
$\int dz (\varepsilon^\pm -\half\psi^\pm\dz x^{9-0})$).

Also, since for each term in
the picture-changing operators,
$(n_\sp +n_\sm -n_\hp -n_\hm)$ is equal to twice the
number of $\dz x^{9-0}$'s minus twice the number of $\dz x^{9+0}$'s
(which must be zero since there are no vertex operators on $G_1$),
the only possible values for the $n$'s which does not
remove the divergence is $n_{\xi^\pm}=n_b -1=n_l
=n_{\psi^\pm} +1=
n_{h^\pm} +1=0$.
So the resulting operator at $p_1$ is
$$|c e^{-\phi^+ -\phi^-} e^{\hp+\hm}\sp\sm|^2,\eqno(III.9)$$
which can be associated with the target-space dilaton in the
$(-1,-1)$ ghost picure since $e^{\hp+\hm} \sp\sm$ is precisely
the matter part of the screening charge that couples to the two-dimensional
curvature.

Now suppose $R$ is either $B_1$ or one of the $H_i$'s, so $g_1=1$.
Changing the location of a picture-changing operator, $F^\pm$,
replaces it with the integral of $[Q,\dz\xi^\pm]$, and after pulling
$Q$ through the Beltrami differential for $K_i$ (this total
derivative is harmless since it has already been shown that
the amplitude is finite near $K_i=0$), $Q$ is left surrounding
the operator at the point $p_1$. Since
$$[Q,
c e^{-\phi^+ -\phi^-} e^{\hp+\hm}\sp\sm]=c(\eta^+ e^{-\phi^-}e^{\hm}\sp
+\eta^- e^{-\phi^+}e^{\hp}\sm),\eqno(III.10)$$
there are no terms with zero modes of both $\sp$ and $\sm$, and therefore
$A_1$ is independent of the locations of the picture-changing operators.
So all of the picture-changing operators can be moved to $p_1$, forcing
the zero modes of $\psi^\pm$ to be cancelled and $A_1$ to vanish
(the $\psi^\pm$ path integral becomes proportional to
$\Theta ([(g_1 -1)p_1 -\Delta_1 ])$, which vanishes by the Riemann
identity).

To prove $A_1$ is zero for the other $B_i$ radii, use precisely the
same argument inductively in $i$ (the total derivative one gets
when calculating $A_1$ near $B_i=0$ is harmless once $A$ has been
shown to be finite near $B_{i-1}=0$).

So the Green-Schwarz superstring amplitudes of equation (II.7) have been
shown to be divergence-free near the limiting points of the
Teichmuller parameters, and are therefore finite. Furthermore, it
was proven in reference 17 that these amplitudes satisfy the
non-renormalization theorem, i.e. all multiloop
amplitudes with fewer than four external massless states vanish (this
proof relied on the assumption that total derivatives coming
from BRST-trivial operators don't contribute to scattering amplitudes,
which has now been verified since finiteness implies that there
is no need to introduce a cutoff in the moduli space).

\vskip 12pt
\centerline{\bf IV. Unitarity}
\vskip 12 pt

In this section, it will be proven that the scattering amplitude
of equation (II.7) for external massless boson-boson states
is equivalent to the light-cone Green-Schwarz amplitude, and
is therefore unitary. This proof will not rely on any assumptions
such as those made in reference 17 concerning the contributions of
the non-light-cone parts of the picture-changing operators.

The first step is to choose the usual light-cone moduli
for the surface, $\tilde\rho_a -\tilde\rho_1$ for $a=2$ to
$2g-2+N$ (the complex interaction-point locations), $A_j$ for
$j=1$ to $g$ (the light-cone momenta of the internal loops),
and $B_j$ for $j=1$ to $g$ (the twists when going around
a loop).$^{15}$ These moduli can be understood as coming from the
unique meromorphic one-form, $\dz\rho$, which has poles of
residue $\alpha_r$ at the $N$ punctures ($\alpha_r$ is the
$k^{9+0}$ momentum of the $r^{th}$ external particle) and
purely imaginary periods, $A_j$ and $B_j$,
when integrated around the $j^{th}$ $a$-cycle and $b$-cycle.

The locations of the picture-changing operators
will be chosen such that
$|F^+ F^-|^2$
sits at each of the $2g-2+N$ zeros of the one-form, i.e. at
the $\tilde\rho_a$'s. Note that since there is no cutoff in
the moduli space, there is no contribution from the total
derivatives which arise when changing to this light-cone
parameterization of the surface.

With this choice of the moduli, the Beltrami differentials
contribute$^{26}$
$$|\prod_{a=2}^{2g-2+N} (\int_{C_a} d\rho \hat b(\rho) -
\int_{C_1} d\rho \hat b(\rho))|^2\quad
\prod_{j=1}^g ( \int_{a_j} d\rho \hat b(\rho) -
 \int_{a_j} d\bar\rho \hat {\bar b}(\rho) )
( \int_{b_j} d\rho \hat b(\rho) -
 \int_{b_j} d\bar\rho \hat {\bar b}(\rho) ) ,\eqno(IV.1)$$
where $\hat b \equiv (\dz\rho)^{-2} b$, $C_a$ is a small
circle surrounding $\tilde\rho_a$,
and $a_j$,$b_j$
are the $g$ $a$-cycles and $b$-cycles.

The vertex operators for the external boson-boson states will be chosen
in light-cone gauge, i.e. $p_{9-0, \mu}$ and $p_{\mu, 9-0}$ are gauged
to zero. In this gauge, the bosonic vertex operator with polarization
in the $4_{-\half}$ direction is
$V_{-l}=c e^{-\phi^+ -\phi^-} (\sm e^\hp \Gml -{k^{-l} \over k^{9+0}}
\sp\sm ) e^{ik_\mu x^\mu}$, and the bosonic vertex operator with
polarization in the
$\bar 4_{+\half}$ direction is
$V_{+\bar l}
=c e^{-\phi^+ -\phi^-} (\sp e^\hm \Gplb -{k^{+\bar l} \over k^{9+0}}
\sp\sm ) e^{ik_\mu x^\mu}$.

Since the picture-changing operators must contribute at least
$2g-2+N$ $\varepsilon$'s in order to have a non-zero amplitude, each
factor of $F^+ F^-$ at the interaction points must contribute an
average of one more $\varepsilon$
than $\psi$. For this to happen, each $F^+ F^-$ factor must contribute
either
$$f_+= e^{\phi^+} e^{-\hp} \vep [e^{\phi^-}(\dz x^{-l}\Gplb +
e^\hp \vem \dz x^{9+0} + \gamma^+ b) +c\dz\xp] \quad or \eqno(IV.2)$$
$$f_-= e^{\phi^-} e^{-\hm} \vem [e^{\phi^+}(\dz x^{+\bar l}\Gml +
e^{-\hp} \sp \dz x^{9-0} + \gamma^- b) +c\dz\xm],$$
where normal-ordering needs to be defined for the terms in $f_\pm$
that involve $\gamma^\pm b$.

Normal-ordering will be defined by :$F^+ F^-$: =$\{Q,\xp \{Q,\xm\}\}$,
which corresponds to first integrating over the fermionic moduli
that couple to the fermionic stress-energy tensor $G^-$, and then
integrating over the fermionic moduli
that couple to $G^+$. With this definition, the term
:$(e^{\phi^-} b\gamma^+)(e^{\phi^+} e^{-\hp}\vep)$: in $f_+$
becomes
$$\dz(b e^{\phi^-}\eta^+) e^{2\phi^+}e^{-\hp}\vep +\half
b e^{\phi^-}\eta^+ e^{-\hp}\vep \dz (e^{2\phi^+}), \eqno(IV.3)$$
while the term,
:$(e^{\phi^-} e^{-\hm}\vem)(e^{\phi^+} b\gamma^-)$: in $f_-$
becomes
$$\dz (e^{-\hm}\vem) e^{2\phi^-}b\eta^- e^{\phi^+} +\half
b e^{\phi^+}\eta^- e^{-\hm}\vem \dz (e^{2\phi^-}). \eqno(IV.4)$$

Note that each $\dz x^{9+0}$ factor in $f_+$ must be balanced with
a $\dz x^{9-0}$ factor from $f_-$ since $\dz\rho|_{\tilde\rho_a} =0$
implies that contracting $\dz x^{9+0}$ with the external momentum
factors, $e^{i\alpha_r k_r^{9-0}}$, gives zero. Also, all $v$ ghosts
can be safely ignored since there are no $u$ ghosts to cancel them,
and all $\sp\sm$ factors coming from the vertex operators can
be ignored since there are no extra $\varepsilon$ factors to absorb
them.

It is convenient to attach an instanton-number-changing operator,
$I^{-1}$, to each of
the $m$ vertex operators of polarization $\bar 4_{+\half}$,
and to attach $g-1+m$ instanton-number-changing operators, $I^{+1}$,
to the interaction-point operators of type $f_-$ (since the integrand
of the amplitude, rather than just the integral, is independent
of the locations of the $I$'s, there is no problem with moving the
$I$'s to different locations for different splittings of the $2g-2+N$
$F^+ F^-$'s into $f_+$'s and $f_-$'s).
As was discussed in reference 17,
this addition of a total instanton
number of $g-1$
shifts the conformal weights of all U(1) transforming fields
by half of their U(1) charge, so the fields $[\Gplb,\Gml,\beta^+,
\beta^-, \gamma^+,\gamma^-, e^\hp, e^\hm]$ now have conformal
weights $[1,0,2,1,0,-1,1,0]$.

It is easy to check
that in order to get a non-zero amplitude,
there must be precisely $(g-1+m)$ $f_-$ operators.
So since $h^-$ no longer appears anywhere in the integrand, the path
integral over the $(\hp,\hm)$ fields is trivial, giving a factor
of $(Z_1(\tau))^{-1}$ that cancels the $Z_1 (\tau)$ factor coming
from the path integral over the $(u,v)$ ghosts.

After combining with the Beltrami differential contribution at
$\tilde\rho_a$ from the loop $C_a$, the operators at the interaction
points become:

$$\int_{C_a} d\rho \hat b f_+=
({\partial^2\rho \over dz^2})^{-1}
[b e^{\phi^+ +\phi^-} \vep
(\dz x^{-l}\Gplb +
\vem \dz x^{9+0})\eqno(IV.5)$$
$$+\vep e^{\phi^+}\dz\xp (:bc:  -\half
({\partial^2\rho \over dz^2})^{-1}
{\partial^3\rho \over dz^3} )
+b\dz b \vep e^{\phi^- +2\phi^+}\eta^+]$$
$$=(be^{\phi^+ +\phi^-}\vep) \int
d\tilde\kappa_a \exp [\lim_{\rho\to\tilde\rho_a}
(\rho -\tilde\rho_a) \tilde\kappa_a
(\drho x^{-l}\hat\Gamma^{+\bar l} +
\hat\varepsilon^-
\drho x^{9+0} +\hat b \gamma^+ +\hat\beta^+\drho \hat c)], $$

$$\int_{C_a} d\rho \hat b I f_-=
({\partial^2\rho \over dz^2})^{-1}
\epsilon^{lmnp} [b e^{2\phi^-} (\vem
\dz x^{+\bar l}\Gpmb\Gpnb\Gppb +
\Gplb\Gpmb\Gpnb\Gppb\dz x^{9-0}) \eqno(IV.6)$$
$$+\Gplb\Gpmb\Gpnb\Gppb \vem e^{2\phi^-}\dz\xm
(:bc:  -\half
({\partial^2\rho \over dz^2})^{-1}
{\partial^3\rho \over dz^3} )]$$
$$= ( ({\partial^2\rho \over dz^2})^{-1}
\epsilon^{lmnp}
\Gplb\Gpmb\Gpnb\Gppb
be^{2\phi^-}\vem) \int
d\tilde\kappa_a \exp [\lim_{\rho\to\tilde\rho_a}
(\rho -\tilde\rho_a) \tilde\kappa_a
(\drho x^{+\bar l}\Gml +
\sp \drho x^{9-0} +\hat\beta^-\drho \hat c)], $$
where $\hat\Gamma^{+\bar l}
=(\dz\rho)^{-1}\Gplb$, $\hat\varepsilon^\pm=
(\dz\rho)^{-1}\varepsilon^\pm$, $\hat\beta^+=(\dz\rho)^{-2}\beta^+$,
$\hat\beta^-=(\dz\rho)^{-1}\beta^-$,
$\hat\gamma^+=\gamma^+$,
$\hat\gamma^-=(\dz\rho)^{+1}\gamma^-$, and $\hat c=(\dz\rho)^{+1} c$.

Note that the $\hat\beta^+\drho\hat c$ term in $f_+$ can be dropped
since there is no $\gamma^-$ term in $f_-$ to absorb it. After performing
the shift of variables, $x^{9\pm 0}\to x^{9\pm 0}-\sum_{r=1}^N
k_r^{9\pm 0} n(z,z_r)$, where $n$ is the Neumann function on the surface,
the $\exp ({ik_r^{9+0} x^{9-0}+
ik_r^{9-0} x^{9+0}})$ factors in the vertex operators
are replaced by $\exp(\sum_{r\neq s} k_r^{9-0} k_s^{9+0} n(z_r,z_s))$.
This shift leaves the interaction-point operators unchanged
since $\dz\delta x^{9+0}|_{\tilde\rho_a}$ is zero and there are no
extra $\vem$'s available to absorb a $\sp$ that is unaccompanied by
a $\dz x^{9-0}$ (i.e., all extra $\vem$'s are accompanied by $\dz x^{9+0}$).

Once the interaction-point operators and vertex operators are in this
form, it can easily be shown that the path integrals over the
non-light-cone matter fields precisely cancel the path integrals over
the ghost fields. This is done by pairing each non-light-cone
matter field with a ghost field in the following way:
$$(x^{9-0},\hat c +\hat{\bar c}),\quad
(\drho x^{9+0},\hat b),\quad
(\drhobar x^{9+0},\hat{\bar b}),\quad
(\psi^+, \hat\beta^- ),\quad
(\psi^-, \hat\beta^+ ),\quad
(\hat\varepsilon^+, \hat\gamma^-),\quad
(\hat\varepsilon^-, \hat\gamma^+)\eqno(IV.7)$$
(this technique can also be used in the Neveu-Schwarz-Ramond
formalism, in which case $\hat\Gamma^{9+0}$ is paired with $\hat\beta$
and $\hat\Gamma^{9-0}$ is paired with $\hat\gamma$). Because of the
Beltrami differential contributions of
$[\prod_{j=1}^g ( \int_{a_j} d\rho \hat b(\rho) -
 \int_{a_j} d\bar\rho \hat {\bar b}(\rho) )
( \int_{b_j} d\rho \hat b(\rho) -
 \int_{b_j} d\bar\rho \hat {\bar b}(\rho) )]$,
$\hat b$ and $\hat {\bar b}$ have the same periodicity
conditions as $\drho x^{9-0}$ and $\drhobar x^{9-0}$. Since the
zero modes of $x^{9\pm 0}$ are absent due to momentum conservation,
the zero modes of $\hat c$ and $\hat{\bar c}$ must also be
absent (removing these zero modes is the same as replacing the
Beltrami differential contribution,
$|\prod_{a=2}^{2g-2+N} (\int_{C_a} d\rho \hat b(\rho) -
\int_{C_1} d\rho \hat b(\rho))|^2$ with
$|\prod_{a=1}^{2g-2+N} \int_{C_a} d\rho \hat b(\rho)|^2$).

It is easily checked that these matched pairs of non-light-cone
matter fields and ghost fields have the same boundary conditions
at all of the interaction points and vertex operators. For
example, ignoring the $\tilde\kappa_a$ dependence, the $\sp$
and $\hat\beta^-$ fields behave like $z^{-1}$ near $f_-$
interaction points, like $z$ near vertex operators of
polarization $\bar 4_{+\half}$, and are regular everywhere else.
Since the
$\tilde\kappa_a$ dependence of these matched pairs is also
identical (i.e., after dropping the $\hat\beta^+
\drho\hat c$ term, the matched pairs occur in the
same quadratic combinations in the exponential term), and since
each matched pair consists of one boson and one fermion, the path
integrals for the matched pairs precisely cancel each other.

After performing these path integrals, only the light-cone matter
fields remain and it is easy to check that the remaining parts of the
vertex and interaction-point operators are precisely the light-cone
Green-Schwarz vertex and interaction-point operators (when
SO(8) is broken down to SU(4)xU(1) in such a way that the SO(8)
vector splits into $4_{-\half}$ and $\bar 4_{+\half}$ representations,
and when boundary conditions on $\Gplb$ and $\Gml$ are chosen to
correspond to those of fields with conformal weight $+1$ and $0$,$^{12}$
the light-cone Green-Schwarz interaction-point operator is
simply $f_+ +f_-$)$^{27}$.

So the two different formalisms give the same scattering amplitudes,
thereby proving the unitarity of the Lorentz-invariant amplitudes
of equation (II.7). Note that this comparison of scattering amplitudes
was only done for light-cone diagrams that did not contain
colliding interaction points (it was assumed that $\dsr$
is non-zero at the interaction points). However, since the
form of the light-cone contact term is completely determined
by the condition that the amplitudes are divergence-free
when two or more interaction points collide, and since the
amplitudes of equation (II.7) contain no such divergences (the
picture-changing operators need not sit on the interaction points),
the two formalisms must also give equivalent amplitudes for
light-cone diagrams that contain colliding interaction points.

\vskip 12pt
\centerline{\bf V. Conclusion}
\vskip 12 pt

In this paper, it was shown that the previously calculated Lorentz-invariant
Type IIB
Green-Schwarz superstring amplitudes
for external massless boson-boson states contain
no unphysical divergences, and are therefore finite.
Furthermore, it was shown that these amplitudes are unitary
since they agree with amplitudes obtained using the light-cone
Green-Schwarz formalism (this also proves the finiteness and
Lorentz invariance of the light-cone Green-Schwarz formalism).

These proofs required the use of external massless boson-boson
states since the other massless states can not be written in a form
with zero instanton charge. Nevertheless, it should be possible
to generalize the proofs of finiteness and unitarity for amplitudes
involving massless fermionic states which carry non-zero instanton charge.
It should also be straightforward to construct BRST-invariant
vertex operators for the massive particles out of the covariant
$x_\pm^\mu$ and $\theta^\alpha_\pm$ fields, and use them to
calculate Lorentz-invariant amplitudes involving external massive
states. Although these should be finite, the two and three-point
amplitudes are no longer expected to vanish.

One disadvantage of the Lorentz-covariant amplitudes of equation (II.7)
is that they are not manifestly N=2 worldsheet supersymmetric. Since
the fermionic stress-energy tensor is not a quadratic function of the
longtitudinal matter fields, it is not obvious how to combine these
matter fields into N=2 superfields. Although making the N=2
worldsheet supersymmetry manifest is not necessary, it would
be nice to express the Green-Schwarz superstring amplitudes as
super-integrals over N=2 bosonic and fermionic moduli.

\vskip 24pt
\centerline {\bf Acknowledgements}
\vskip 12pt
I would like to thank P.DiVecchia, M.Freeman, M.Green,
P.Howe, S. Mandelstam, J.Petersen, F. Pezzella,
A.Restuccia, J.Sidenius, J.Taylor,
A.Tollsten, and P.West
for useful discussions, and the SERC for its financial support.

\vskip 24pt

\centerline{\bf References}
\vskip 12pt

\item{(1)} Mandelstam,S., Nucl.Phys.B69 (1974), p.77.

\item{(2)} Greensite,J. and Klinkhamer,F.R., Nucl.Phys.B291 (1987), p.557.

\item{(3)} Mandelstam,S., private communication.

\item{(4)} Friedan,D., Martinec,E., and Shenker,S., Nucl.Phys.B271
(1986), p.93.

\item{(5)} Verlinde,E. and Verlinde,H., Phys.Lett.B192 (1987), p.95.

\item{(6)} Atick,J. and Sen,A., Nucl.Phys.B296 (1988), p.157.

\item{(7)} Mandelstam,S., ``The n-loop Amplitude: Explicit Formulas,
Finiteness, and Absence of Ambiguities'', preprint UCB-PTH-91/53,
October 1991.

\item{(8)} Berkovits,N., Nucl.Phys.B304 (1988), p.537.

\item{(9)} Aoki,K., D'Hoker,E., and Phong,D.H., Nucl.Phys.B342 (1990),
p.149.

\item{(10)} Green,M.B. and Schwarz,J.H., Nucl.Phys.B243 (1984), p.475.

\item{(11)} Mandelstam,S., Prog.Theor.Phys.Suppl.86 (1986), p.163.

\item{(12)} Restuccia,A. and Taylor,J.G., Phys.Rep.174 (1989), p.283.

\item{(13)} Carlip,S., Nucl.Phys.B284 (1987), p.365.

\item{(14)} Gilbert,G. and Johnston,D., Phys.Lett.B205 (1988), p.273.

\item{(15)} Mandelstam,S., ``The interacting-string
picture and functional integration''
in 1985 Santa Barbara Workshop on Unified
String Theories, eds. M.B. Green and D. Gross (World Scientific,
Singapore), p.577.

\item{(16)} Kallosh,R. and Morosov,A., Phys.Lett.B207 (1988), p.164.

\item{(17)} Berkovits,N., ``Calculation of Greeen-Schwarz Superstring
Amplitudes using the N=2 Twistor-String Formalism'', SUNY at Stonybrook
preprint ITP-SB-92-42, August 1992, to appear in
Nucl.Phys.B, hep-th bulletin board 9208035.

\item{(18)} Berkovits,N., Phys.Lett.B300 (1993), p.53, hep-th
bulletin board 9211025.

\item{(19)} Ademollo,M., Brink,L., D'Adda,A., D'Auria,R.,
Napolitano,E., Sciuto,S., Del Giudice,E., DiVecchia,P., Ferrara,S.,
Gliozzi, F., Musto,R., Pettorini,R., and Schwarz,J., Nucl.Phys.B111
(1976), p.77.

\item{(20)} Rocek,M., private communication.

\item{(21)} Verlinde,E. and Verlinde,H., Nucl.Phys.B288 (1987), p.357.

\item{(22)} Lechtenfeld,O., Phys.Lett.B232 (1989), p.193.

\item{(23)} Petersen,J.L. and Sidenius,J.R., Nucl.Phys.B301 (1988), p.247.

\item{(24)} Martinec,E., Phys.Lett.B171 (1986), p.189.

\item{(25)} DiVechhia,P., Pezzella,F., Frau,M., Hornfeck,K.,
Lerda,A., and Sciuto,S., Nucl.Phys.B322 (1989), p.317.

\item{(26)} D'Hoker,E. and Giddings,S., Nucl.Phys.B291 (1987), p.90.

\item{(27)} Berkovits,N., Nucl.Phys.B379 (1992), p.96., hep-th bulletin
board 9201004.

\end